\documentclass[manuscript]{emulateapj}
\usepackage{color}
\usepackage{hyperref}

\shorttitle{A Global Correction to PPMXL Proper Motions}
\shortauthors{Vickers et al.}

\begin{document}

\title{A Global Correction to PPMXL Proper Motions}

\author{John J. Vickers\altaffilmark{1,2,$\dagger$}, Siegfried R\"{o}ser\altaffilmark{1}, Eva K. Grebel\altaffilmark{1}}

\altaffiltext{1}{Astronomisches Rechen-Institut, Zentrum f\"{u}r Astronomie der Universit\"{a}t Heidelberg, M\"{o}nchhofstr. 12-14, 69120 Heidelberg, Germany}
\altaffiltext{2}{Key Laboratory for Research in Galaxies and Cosmology, Shanghai Astronomical Observatory, Chinese Academy of Sciences, 80 Nandan Road, Shanghai 200030, China}

\altaffiltext{$\dagger$}{johnjvickers@shao.ac.cn}

\begin{abstract}
In this paper we notice that extragalactic sources seem to have non-zero proper motions in the PPMXL proper motion catalog. We collect a large, all-sky sample of extragalactic objects and fit their reported PPMXL proper motions to an ensemble of spherical harmonics in magnitude shells. A magnitude dependent proper motion correction is thus constructed.

This correction is applied to a set of fundamental radio sources, quasars, and is compared to similar corrections to assess its utility. We publish, along with this paper, code\footnote{available at time of publication of this preprint at: \url{https://github.com/johnjvickers/ppmxl_correction}} which may be used to correct proper motions in the PPMXL catalog over the full sky which have 2 Micron All Sky Survey photometry.
\end{abstract}

\section{Introduction}\label{sec:introduction}
Proper motions are perhaps one of the most sought after astrometric measurements in modern astronomy, along with precision distance determinations from astrometric parallax; the two of which, when combined with radial velocity, unlock so-called `6D-phase space' studies which have the potential to discover numerous velocity substructures. In measuring the motion of astronomical sources in the plane of the sky, a constant tradeoff must be made between the time scale over which to measure the motion (with longer time scales allowing for more accurate estimation of the motion) and the precision of the data used in the estimate (with modern telescopes and CCDs being much more precise than older ones).  Another property of proper motions is that their most commonly used derivative, transverse velocities, suffer a problem similar to photometric distance estimations in that their errors are a function of the distance to the source.

However, proper motions are critical in the understanding of the kinematics of systems such as globular clusters (\citealt{cas2013} and the others in that series), streams (\citealt{kop2013}, \citealt{din2005}), open clusters \citep{van2009}, nearby cool dwarf stars \citep{dea2009} and the Galaxy as a whole \citep{kui2002}. They provide a wealth of phase space information and will only become more powerful with the precision astrometry and parallax measurements promised by the Gaia satellite. Therefore it is imperative that much care and effort be taken to ensure that proper motion measurements are comparable to each other, and are on an inertial frame.

The PPMXL proper motion catalog \citep{roe2010} is a database of $\sim$900 million proper motion estimates calibrated to the International Celestial Reference System (ICRS\footnote{The International Celestial Reference System is defined as an inertial reference system whose pole is ``close" to the pole of FK5 at equinox and epoch 2000.0 and the zeropoint on the equator ``close" to the zero-point of FK5}, \citealt{fey2009}) for objects across the entire sky down to a magnitude of V$\approx$20. The catalog was created by combining United States Naval Observatory B (USNO-B1.0; \citealt{mon2003}) catalog astrometry with 2 Micron All Sky Survey (2MASS; \citealt{skr2006}) astrometry and bringing that relative astrometry into the ICRS by comparison with the Hipparcos catalog \citep{esa1997}, which is the primary ICRS reference catalog.

\begin{figure}[h!]
\includegraphics[width=\linewidth]{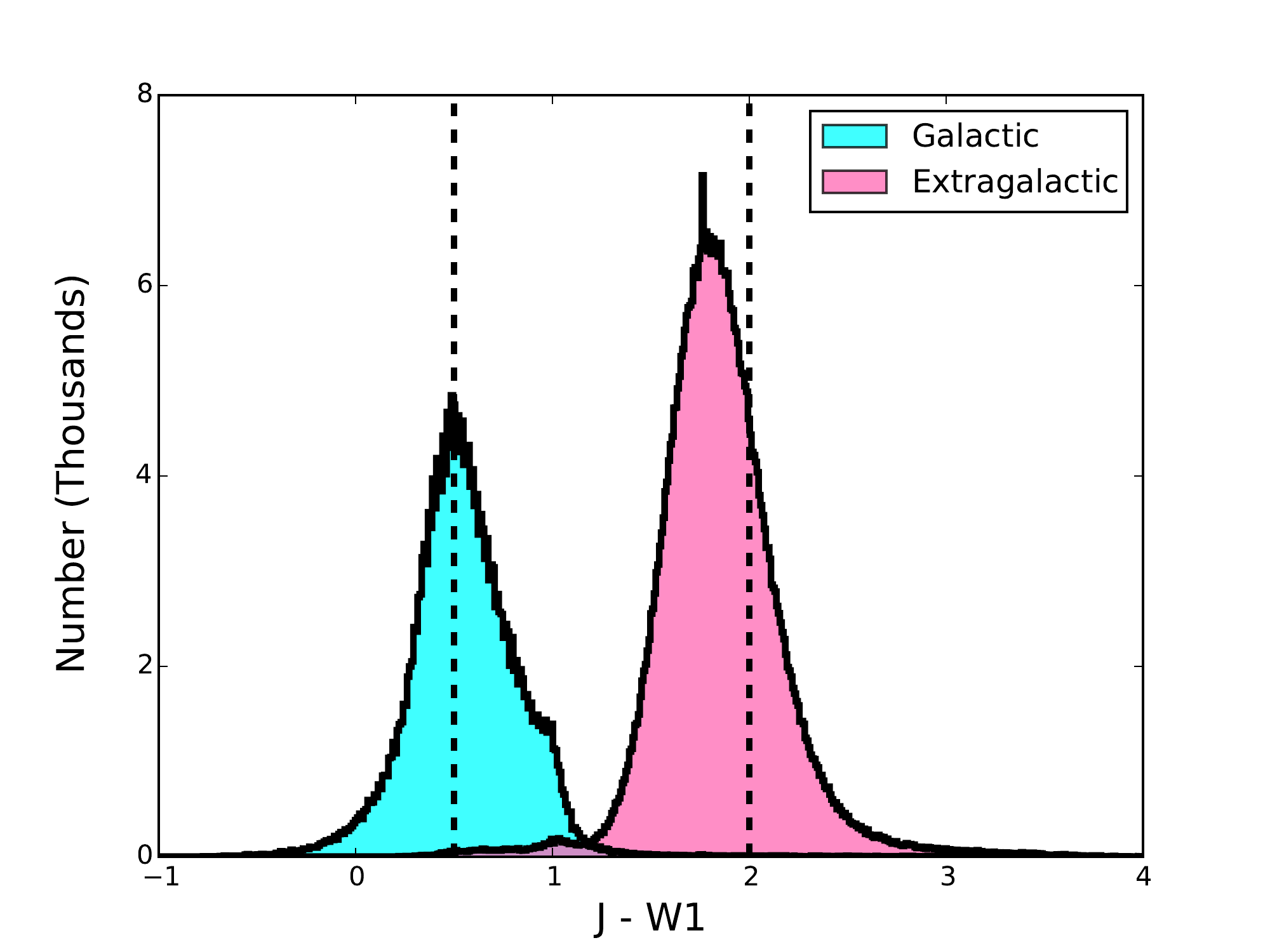}
\caption{A color selection for Galactic (cyan) vs extragalactic (pink) objects (verified by SDSS spectroscopy) using WISE and 2MASS photometry. The vertical line at J-W1 = 2.0 indicates the color cut we use to select extragalactic objects; redward of this line, $\sim$99\% of the data are extragalactic sources. The vertical line at J-W1 = 0.5 is the color cut we use to select stellar objects; blueward of this line, $\sim$99\% of the data are stars (see though, Sections \ref{sec:data} and \ref{sec:eff_plane}). Data selected via these color cuts are plotted spatially in Figure \ref{fig:star_vs_gal_sph_hammer}.}
\label{fig:color_cut}
\end{figure}

\begin{figure*}[t]
\includegraphics[width=\linewidth]{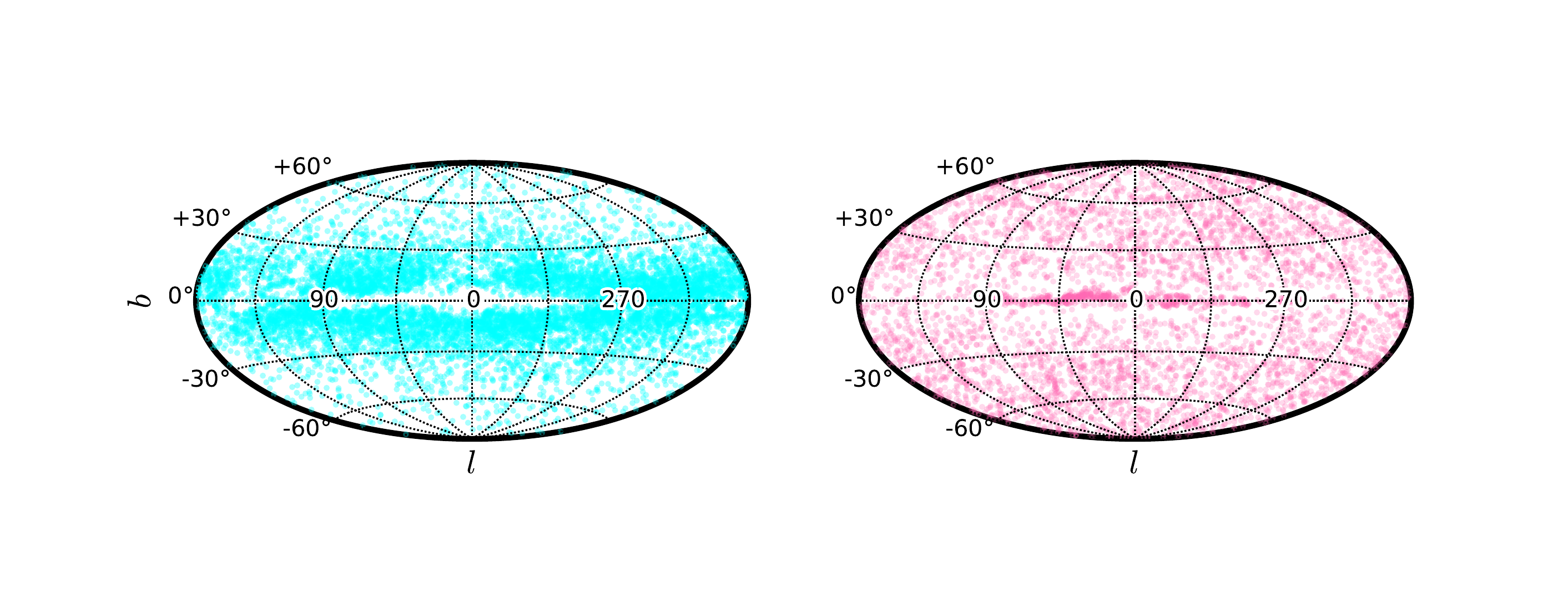}
\caption{Hammer projection of the Galactic coordinate distribution of stars (left) and extragalactic objects (right) as selected by the color cut in Figure \ref{fig:color_cut}. The uniform distribution of the extragalactic objects across the sky as compared to the disk-like distribution of the stars is indicative that the color selection is effective for most of the sky. Extreme reddening at absolute latitudes below 10$^{\circ}$, however, incorrectly flags Galactic stars as extragalactic sources. Note that these figures show a randomly selected one one-hundredth of the data used in the proper motion fitting shown in Figures \ref{fig:radec_gals} and \ref{fig:radec_stars}.}
\label{fig:star_vs_gal_sph_hammer}
\end{figure*}

In general, two methods exist for creating an inertial proper motion catalog: \\

\begin{enumerate}
	\item Using the Hipparcos catalog, which is the primary realization of the ICRS at optical wavelengths, and extending it to fainter magnitudes. This is the approach utilized in the creation of:

	\begin{enumerate}
		\item the Tycho-2 catalog \citep{hog2000}, which is based on Hipparcos astrometry combined with ground observations,
		\item the PPMXL catalog (the details of which are described in \citealt{roe2010}), which uses PPMX \citep{roe2008}, 2MASS and USNO-B1.0 data,
		\item the UCAC4 catalog, which utilizes Hipparcos, Tycho-2, 2MASS, and USNO data (see \citealt{zac2013}).
	\end{enumerate}
	
	\item Using extragalactic sources by forcing their proper motions to zero (since extragalactic sources would have negligible proper motions) and 		referring the positions of these extragalactic sources to the positions of the defining sources of the International Celestial Reference Frame 		(ICRF\footnote{The ICRF realizes an ideal reference system, the ICRS, by precise equatorial coordinates of extragalactic radio sources 			observed in Very Long Baseline Interferometry programs. The Hipparcos catalog which includes all the FK5 stars was astrometrically 			aligned to the ICRF and provides the primary realization of the ICRS at optical wavelengths.}, \citealt{fey2009}). The latter step introduces 		another small rotation onto the proper motion system of stars. This is the approach used by:

	\begin{enumerate}
		\item \citet{mun2004} to create a proper motion catalog for the Sloan Digital Sky Survey (SDSS; \citealt{yor2000}),
		\item \citet{fed2009} to create the XPM proper motion catalogue by combining 2MASS and United States Naval Observatory A 			(USNO-A2.0; \citealt{mon1998}) data.
	\end{enumerate}
\end{enumerate}

Ideally, these two methods will create agreeing catalogs. But relying on the very bright and sparse Hipparcos reference stars one encounters the problem of systematic, magnitude dependent errors in photographic and CCD astrometry. In the construction of PPMXL, \citet{roe2010} could not use the Hipparcos catalog directly, but had to go via intermediate steps to reduce the faint stars and extragalactic objects to the Hipparcos reference frame. The approach presented in this paper allows us to detect systematic errors in the original PPMXL reduction by direct comparison with extragalactic sources. We note in passing that the opposite effect may be found if one were to compare the \citet{mun2004}, catalog at its bright end with the Hipparcos reference frame.

Of the catalogs mentioned, both the PPMXL and UCAC4 catalogs are calibrated to the ICRS via Hipparcos data. Despite this, systematic variation exists between them of up to 4 mas yr$^{-1}$. See \citet{vit2015} for an analysis of these differences using a set of spherical harmonic functions. An offset between proper motions estimated by the ICRS based PPMXL catalog and the extragalactic based SDSS catalog has also been noticed by \citet{wu2011}.  On close inspection, it was noted that galaxies and quasars in the PPMXL catalog had non-zero proper motions with small, but noticeable systematics.

\begin{figure*}[h!]
\centering
\includegraphics[height=0.455\textheight]{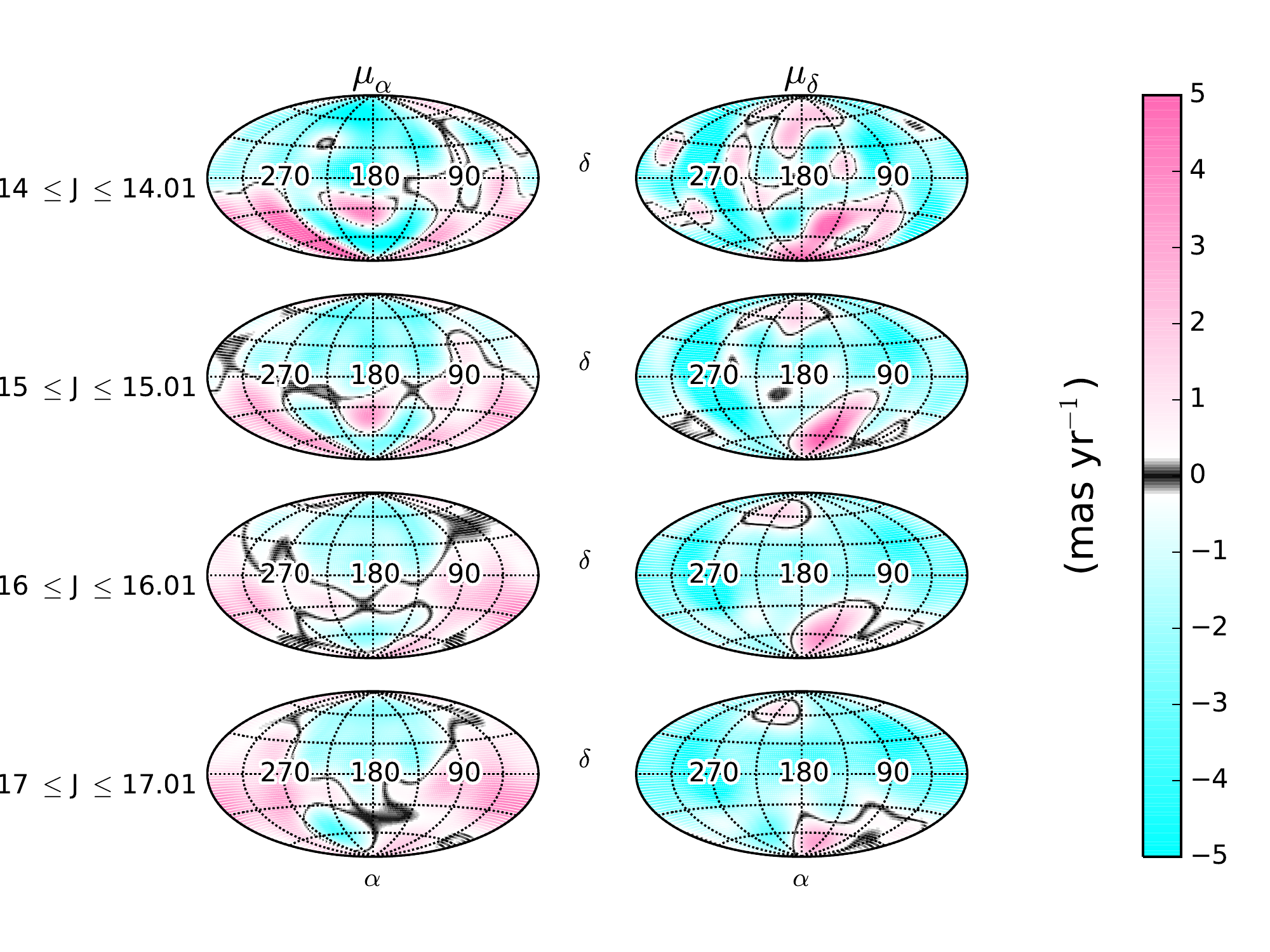}
\caption{Hammer projection of the spherical harmonic fits for our data in the magnitude bins described in Section \ref{sec:data} (the three bins at the half-magnitudes are omitted to save space). There are areas of clear systemic proper motion (for example proper motion in $\delta$ is systematically positive around [$\alpha$, $\delta$] = [145, -60]) and we can see that the fits show some variation with the magnitude slice investigated. Proper motions of our extragalactic data set once this corrective volume has been subtracted are shown in Figure \ref{fig:quiver}.}
\label{fig:radec_gals}
\includegraphics[height=0.455\textheight]{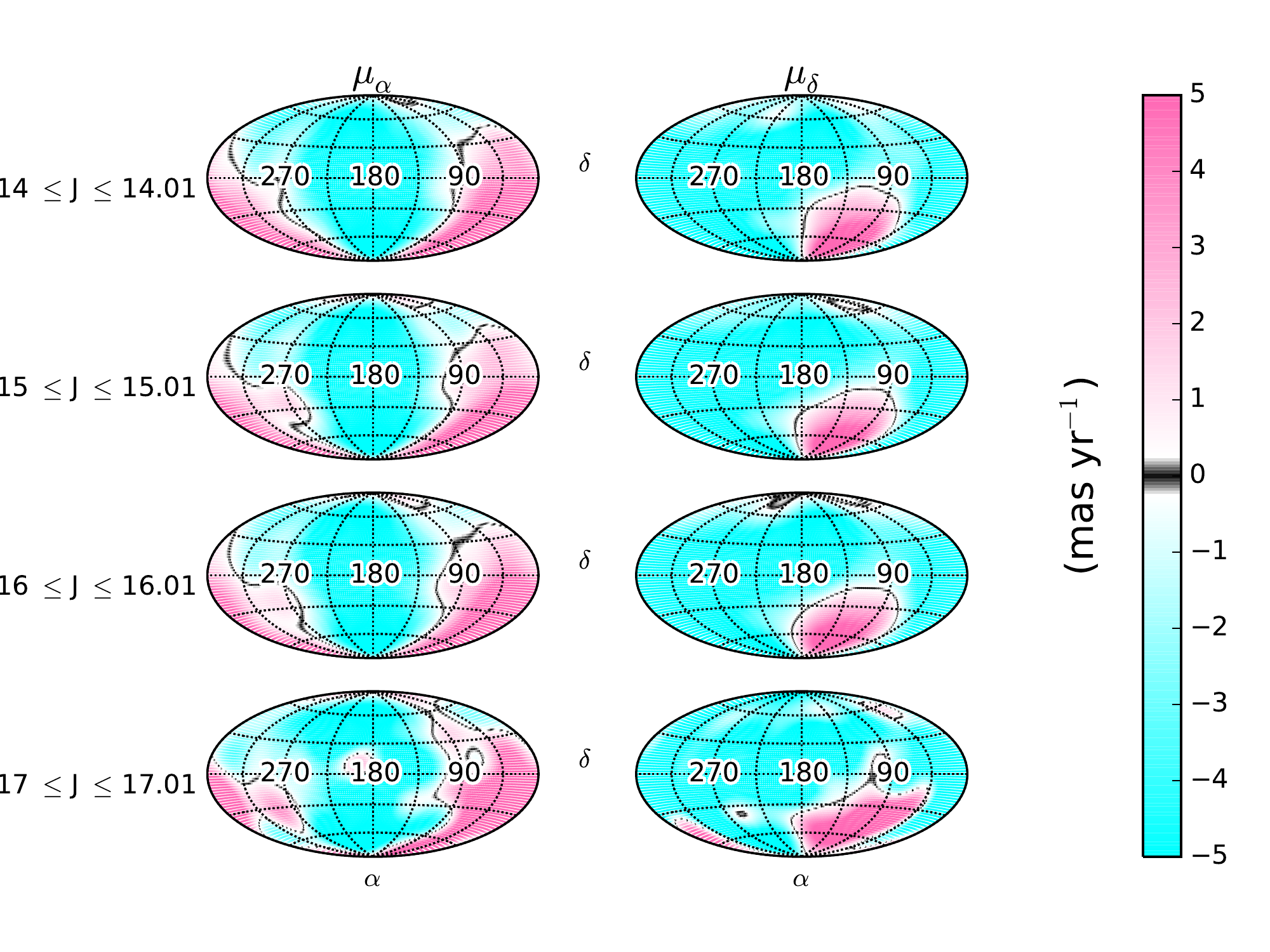}
\caption{Similar to Figure \ref{fig:radec_gals} but instead fit to selected stellar objects. The proper motion profiles of the selected extragalactic objects are different from the proper motions of the stars; further evidence that their movement profiles arise from biases in the PPMXL catalog.}
\label{fig:radec_stars}
\end{figure*}

In this paper, we construct and publish a post-processing corrective script which can be used to correct for this offset in small data sets; we also add two additional columns to the PPMXL database which have been post processed by our method. In Section \ref{sec:data}, we select a sample of extragalactic objects across the whole sky using a combination of Wide-field Infrared Survey Explorer (WISE\footnote{We actually use ALLWISE data, which are a combination of data from the main cryogenic survey of the WISE satellite and the non-cryogenic survey NEOWISE \citep{mai2011} which took place after the main survey.}; \citealt{wri2010}) and 2MASS photometry. In Section \ref{sec:fitting}, we fit spherical surfaces to the PPMXL proper motions of these objects to find local proper motion offsets from zero and create a corrective volume from these surface fits. In Section \ref{sec:testing_the_fits}, we subtract this corrective volume from: the proper motions of our own extragalactic sample, the proper motions of a set of fundamental ICRS radio sources, the proper motions of the quasars of the \citet{ver2010} catalog, and from the independently constructed corrective surface of \citet{gra2014}. These four residuals demonstrate the effectiveness of the correction. Finally, we conclude in Section \ref{sec:discussion}.

\section{Data}\label{sec:data}

To correct the PPMXL proper motion catalogue, we require a sample of extragalactic objects (which are assumed to have intrinsic proper motions of nearly 0 mas yr$^{-1}$). We can use the calculated proper motions of this intrinsically stationary sample to estimate underlying biases in the PPMXL catalog. Ideally the sample we use will cover the whole sky, so that the entire PPMXL catalog can be corrected, and have as many items as possible to maximize the robustness of the fit.

Note that here we are relying on the assumption that extragalactic sources are essentially stationary. This is true when our proper motion precision is on the order of milliarcseconds per year; but in the Gaia era of microarcsecond per year precision \citep{mig2005}, this will not be true. The orbital motion of the Sun around the Galaxy and the Galaxy's own motion in the Local Group will be detectable as an aberration in the movements of extragalactic sources \citep{kov2003}.

At this point we refer to an interesting color selection noted independently by \citet{kov2015} and \citet{pra2014}. Both groups find that extragalactic and Galactic objects separate from each other in a 2MASS J to WISE W1 color (J-W1). This is optimal for our purposes as both WISE and 2MASS are all-sky surveys with a large amount of overlap in their photometric detections.

To investigate how Galactic and extragalactic origin relate to J-W1 color, we initially collect a sample of objects with spectroscopy available from the SDSS. The SDSS has a much higher photometric resolution (FWHM $r\sim$1.3") than WISE (FWHM W1$\sim$6.1") or 2MASS (FWHM J$\sim$2.9"), so it is ideal for identifying galaxies based on their photometric profiles, while the medium resolution spectroscopy (R$\sim$1800) is necessary for separating quasars from stellar objects. These spectroscopic data can thus be split into two groups: Galactic (objects with point spread function photometry \emph{and} star-like spectra); and extragalactic (objects with non-point spread photometry \emph{or} quasar-like spectra) objects.

We then crossmatch these data with the WISE all sky catalog to obtain W1 and 2MASS J magnitudes and calculate J-W1 colors (2MASS photometry is included in the WISE database).

Figure \ref{fig:color_cut} shows a histogram of these two groups in J-W1 space. Redward (blueward) of the vertical line at J-W1 = 2.0 (0.5), defines a sample of objects which are $\sim$99\% extragalactic (Galactic), according to the definitions of extragalactic and Galactic given above. For the rest of this paper, we will use these color cuts as a selector for Galactic and extragalactic objects. The spatial distribution for a subset of these objects is shown in Figure \ref{fig:star_vs_gal_sph_hammer}. Note that these color cuts exclude a majority of the sources in the data set. This is to ensure as pure a selection as possible; completeness is not very important for our purposes.

We note that this estimate is an optimistic estimate and based only on SDSS spectroscopy in low extinction areas of the sky. Since the SDSS spectroscopic survey only sparsely samples the inner regions of the galaxy, it is not immediately apparent how much reddening will affect our selection. However, assuming the filter coefficients of \citet{dav2014} for 2MASS J and WISE W1 bands (calibrated to the SDSS $r$ band coefficient of \citealt{sch2011}), an extinction of E(B-V) $\sim$ 2 will result in a color shift of $\Delta$(J-W1) $\sim$ 1. This is enough to scatter stars from the red wing of the stellar distribution into our extragalactic color selection (see Figure \ref{fig:color_cut}). Thus, areas with extinction equal to or higher than this (absolute Galactic latitudes less than about 10$^{\circ}$) should be treated with caution in this analysis. This is discussed further in Section \ref{sec:eff_plane}

We query the WISE database for data in that red color range in seven, 0.1 magnitude wide, 0.5 magnitude separated, J band slices (ie. J = 14.0-14.1, 14.5-14.6 ... 17.0-17.1). These slice separations are chosen to sample the effective J range of 2MASS and the widths are chosen to select a number of objects which is large compared to the number of parameters in the surface fit ($>$ 10,000 items per slice), while being small enough to query conveniently from the WISE database. These data are crossmatched with the PPMXL proper motion database as a function of position and J magnitude to collect the proper motion data for these objects.

This extragalactic data set will be used to construct an all-sky, magnitude dependent correction for the PPMXL proper motions.

\section{Fitting}\label{sec:fitting}
Our seven data slices should be predominantly composed of extragalactic sources -- if this is true, then the sources should have zero proper motion in any direction (or rather, the proper motions should be so small that they are indistinguishable from zero). This is not the case, we find that these objects have a net movement of 0.26 mas yr$^{-1}$ in $\alpha$ and -2.9 mas yr$^{-1}$ in $\delta$.

We wish to fit these proper motions to a combination of spherical harmonics in each data slice -- such a fitting method removes the discontinuity produced by binning the data at the cost of becoming more computationally expensive than a corrective table. The fit harmonics may then be subtracted from the data to produce a new, recentered PPMXL proper motion catalog.

In each magnitude bin, the first eight degrees of spherical harmonics (the choice of eight degrees is explained in Section \ref{sec:prevention_of_overfitting}) are fit to the proper motions of the data (one fit for the proper motion in $\alpha$ and one fit for the proper motion in $\delta$). So in total we perform 14 fits and each fit has 81 variables; we do not fit inverse harmonics. 

Initially, to remove spurious proper motions from these data, a 2$\sigma$ cut on the proper motions about the mean proper motions in each direction was implemented. However, since the systematic proper motions in the data vary as a function of sky position, a cut of this type will systematically remove more of the high proper motion tail or more of the low proper motion tail for different areas of sky, which can bias the local mean proper motions.

Instead we implement a localized outlier removal procedure. For every object in our fitting data set, its proper motion is compared to the bulk proper motion of the 1000 objects nearest to it on the sky. If its proper motion is inconsistent with the neighborhood bulk proper motion by an amount of more than three times its reported proper motion error, that point is presumed to be spurious and is removed. This cut removes more data than the global 2$\sigma$ cut, but does not shift the local mean proper motions (as the 2$\sigma$ cut does).

The results of this fitting are shown in Figure \ref{fig:radec_gals} -- these surfaces show areas of systemic proper motions in the data. By subtracting this surface from the data, we should essentially re-center these objects on zero proper motion around the celestial sphere. We also fit the motion of stars in a similar manner in Figure \ref{fig:radec_stars}. A comparison shows that the detected motions of the extragalactic objects are not side effects of erroneously identified stars.

We turn these surfaces into a corrective volume by linearly interpolating between the surfaces in magnitude space. In this way we may correct proper motions for objects which do not lie in the magnitude ranges of the data slices we fit.

\begin{figure}[ht!]
\includegraphics[width=\linewidth]{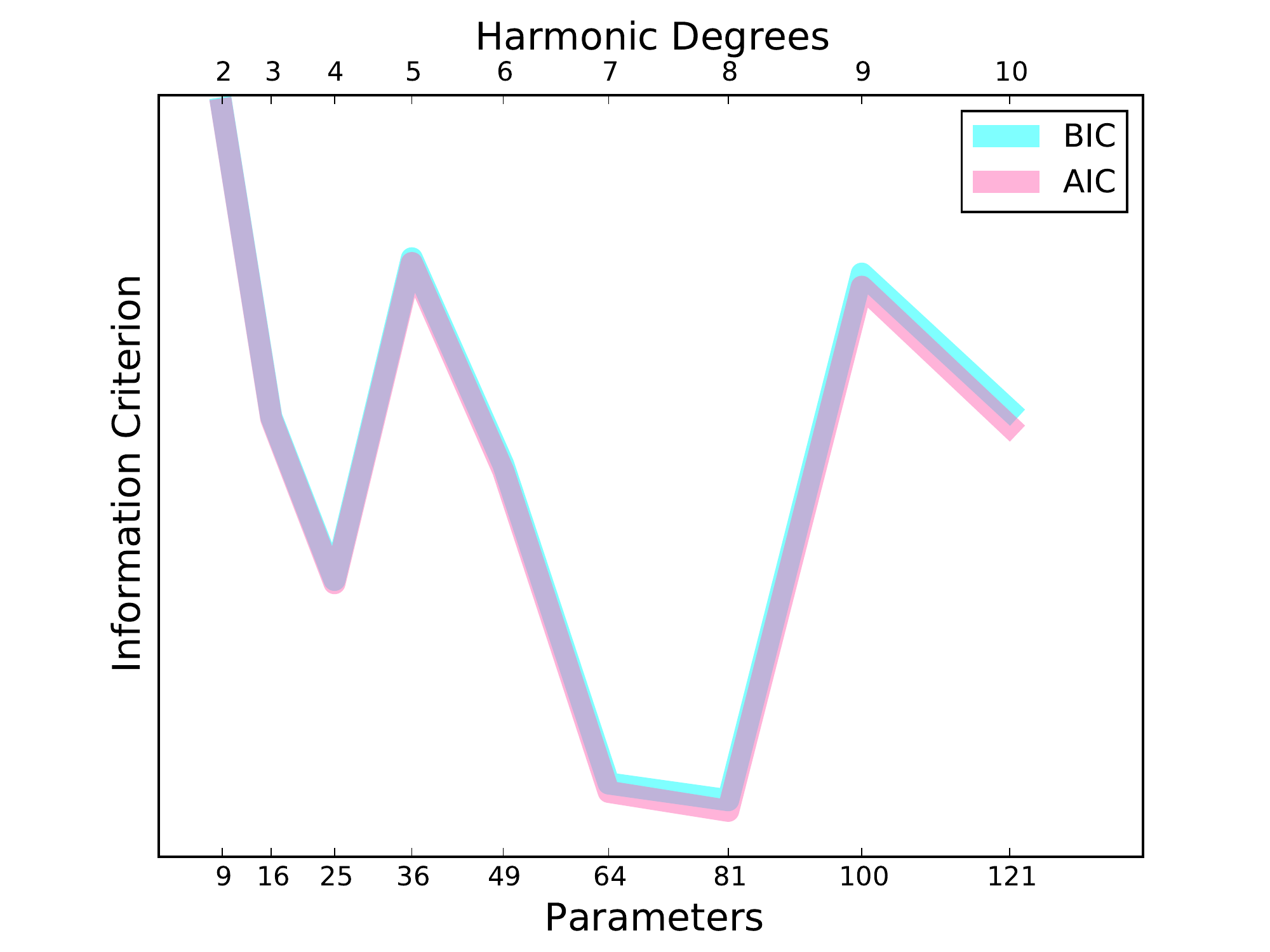}
\caption{Bayesian and Akaike Information Criterion (BIC, AIC) values for spherical harmonic fits to our data. Both tests prefer the 81 parameter, eighth degree spherical harmonic fits.}
\label{fig:bic}
\end{figure}

%HASHTAG FIG
\begin{figure*}[ht!]
\includegraphics[width=\linewidth]{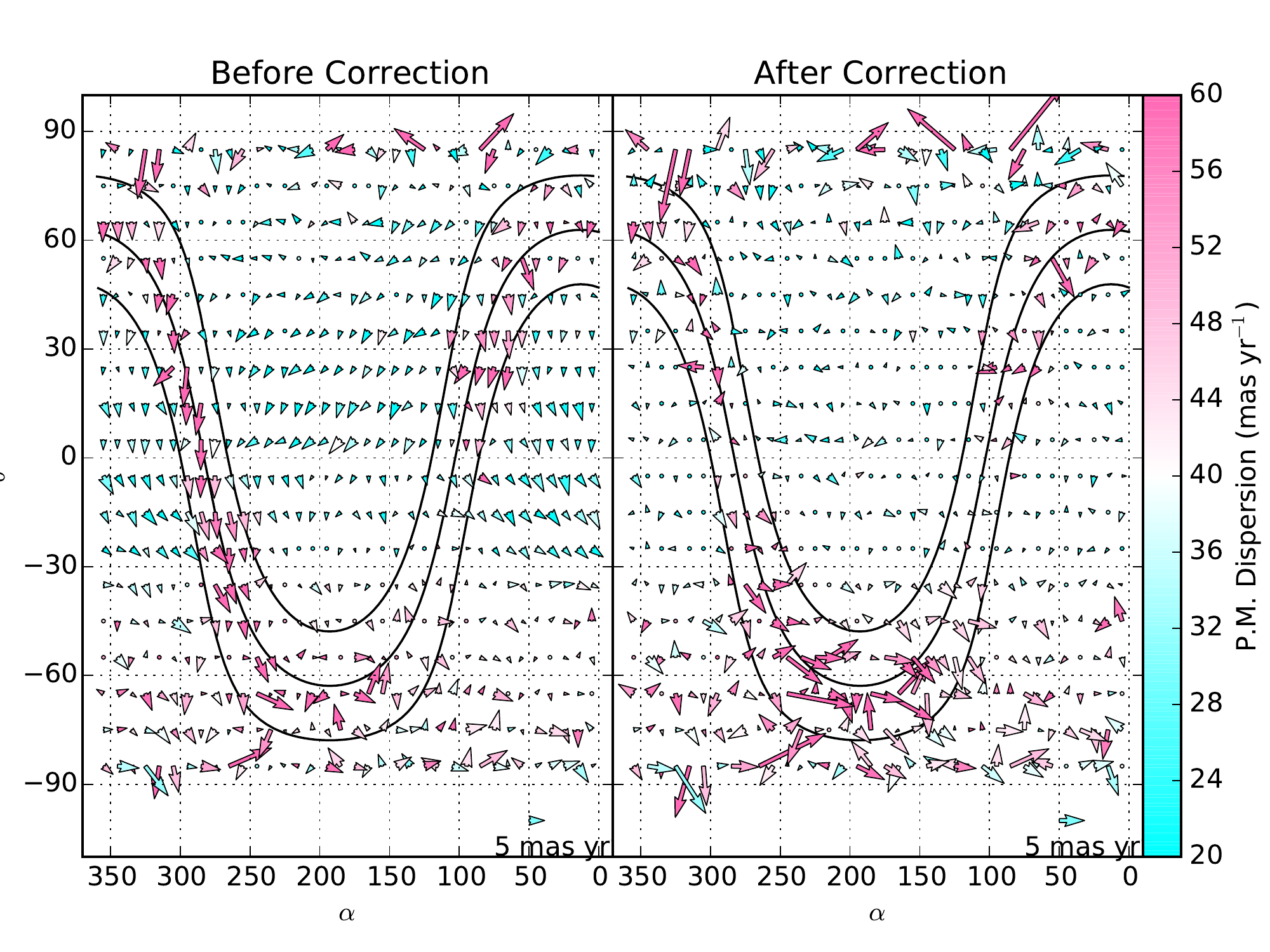}
\caption{Bulk proper motions (each arrow indicates a 10$^{\circ}$x10$^{\circ}$ bin) of our extragalactic sources before (left) and after (right) subtracting the surfaces shown in Figure \ref{fig:radec_gals}. The arrows show the on-sky direction of the bulk proper motions with the length of the arrows indicating the average magnitude of the proper motion (in the bottom right of each panel is a scale-arrow showing 5 mas yr$^{-1}$) and the color of the arrows indicating how dispersed the measurements are within that particular bin. Note that the dispersion appears correlated with the Galactic plane and center -- this indicates imperfect identification of extragalactic sources in these areas. The Galactic plane and $\pm15^{\circ}$ latitude lines are underplotted for reference.}
\label{fig:quiver}
\end{figure*}

\subsection{Prevention of Overfitting and the Choice of an Eighth  Degree Fit}\label{sec:prevention_of_overfitting}

A major consideration when producing a database correction such as this is the computational time involved in implementing the correction. A fit to a tenth degree spherical harmonic as described in Section \ref{sec:fitting} for example has 121 variables, which must then be evaluated for all the points in the given input catalog (over 900 million for the PPMXL catalog). Each degree added to the fit increases the number of parameters dramatically.

To find the optimal fitting degree, we utilize two information criterion tests. An information criterion test is an evaluation of a model fit to a set of data which penalizes models based on the number of parameters and rewards models based on their goodness of fit to the data. Two commonly used information criterion tests are the Akaike Information Criterion and the Bayesian Information Criterion tests.

The Akaike Information Criterion (AIC) test \citep{aka1974} may be described mathematically as:
\begin{equation}
AIC = 2k - 2\mathcal{L},
\label{eqn:aic}
\end{equation}
where $k$ is the number of parameters in the fit, and:
\begin{equation}
\mathcal{L} = \sum log \left( \frac{1}{\sigma_{\mu}\sqrt{2\pi}} e^{\frac{-\mu^{2}}{2\sigma_{\mu}^{2}}} \right),
\label{eqn:likelihood}
\end{equation}
is the likelihood carried out over all objects. $\mu$ and $\sigma_{\mu}$ are the total proper motions and errors on the total proper motions for each individual object, respectively.

The Bayesian Information Criterion (BIC) test \citep{sch1978} is similar, but penalizes higher order models more than the AIC test, and may be described by the equation:
\begin{equation}
BIC = k \cdot log(n) - 2\mathcal{L},
\label{eqn:bic}
\end{equation}
where $n$ is the number of measurements.

In Figure \ref{fig:bic} we show the results of these two tests, the lowest information criterion value indicates the preferred model. Both tests prefer eighth order spherical harmonic fits and form the basis for our choice.

\begin{figure*}[ht!]
\includegraphics[width=\linewidth]{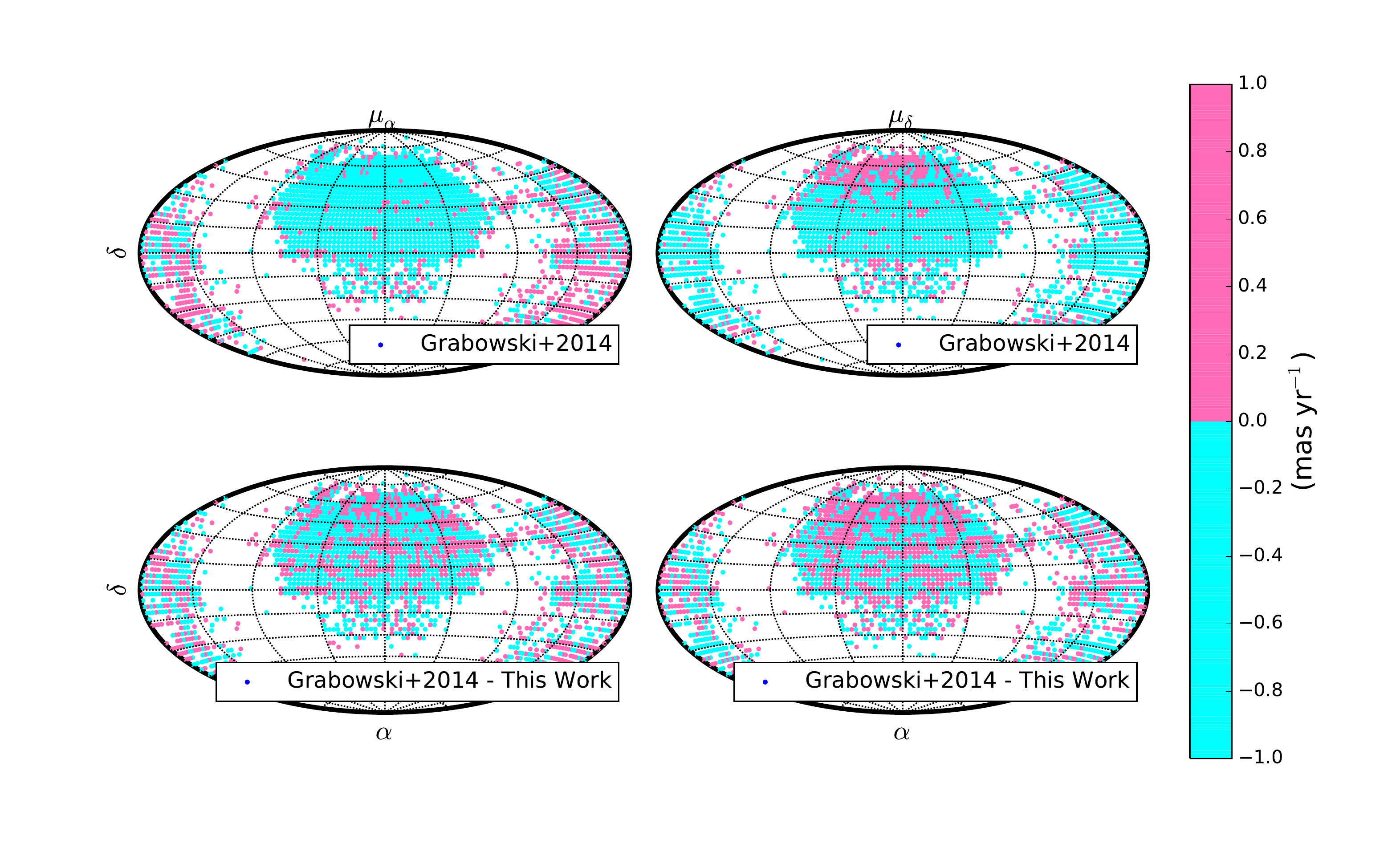}
\caption{The corrective tables for quasar proper motion in the PPMXL in $\alpha$ and $\delta$ (left and right, respectively) presented by \citet{gra2014} (upper panels) and the residuals between their tables and our correction (lower panels). The colors indicate positive and negative proper motion bias, but not the magnitude of the bias. In the upper panels, trends in the PPMXL reported proper motions of quasars lead to hot and cold spots in the catalog footprint, which have been mentioned already. The random appearance of the residuals implies good agreement between our finding and theirs with no overlying, disagreeing trends.}
\label{fig:grabowski}
\end{figure*}

\begin{table*}
\begin{center}
\caption{Efficacy of Recentering}
\begin{tabular}{|c|c|c|c|c|c|c|c|c|}
\tableline
& \multicolumn{4}{|c|}{Average P.M. R.A.} & \multicolumn{4}{|c|}{Average P.M. Dec.} \\
\tableline
& \multicolumn{2}{|c|}{Before} & \multicolumn{2}{|c|}{After} & \multicolumn{2}{|c|}{Before} & \multicolumn{2}{|c|}{After} \\
\tableline
& $\mu$ & $\sigma$ & $\mu$ & $\sigma$ & $\mu$ & $\sigma$ & $\mu$ & $\sigma$ \\
\tableline
WISE & -0.06 & 13.73 & -0.1 & 13.64 & -2.28 & 12.51 & 0.13 & 12.41\\
ICRF VCS & -0.13 & 7.93 & 0.08 & 7.88 & -2.18 & 7.29 & 0.08 & 7.2\\
ICRF Non-VCS & 0.44 & 9.75 & 0.65 & 9.55 & -2.14 & 6.77 & -0.27 & 6.91\\
Veron-Cetty et al. 2010 & -0.82 & 6.92 & -0.07 & 6.69 & -1.78 & 6.23 & 0.3 & 6.13\\
SDSS &  &  & -0.37 & 4.05 &  &  & 0.46 & 3.58\\
\tableline
\end{tabular}
\tablecomments{PPMXL proper motions of various compact radio sources (ICRF VCS, ICRF Non-VCS), quasars (\citealt{ver2010}) and extragalactic sources identified in this paper (WISE) before and after the correction presented in this work. The proper motions of quasars in SDSS are shown for reference.}
\end{center}
\label{tab:efficacy_of_recentering}
\end{table*}

\section{Testing the Fits}\label{sec:testing_the_fits}

\subsection{Internal Consistency}\label{sec:internal_consistency}
We first investigate the success of our methodology in correcting the proper motions of our fitting data set by subtracting the constructed correction volume from the proper motion measurements used to derive it.

Figure \ref{fig:quiver} shows the bulk motions of the data used to construct the correction volume before and after the volume is subtracted from the proper motions. Several trends are noticeable from Figure \ref{fig:quiver}.
\begin{enumerate}
\item The Galactic plane and center are recognizable in their dispersion signatures -- this implies that our color selection is imperfect in these areas.
\item South of $\delta\sim$-45$^{\circ}$ the bulk proper motions appear more random -- this is an artifact of the defining observations used to calibrate the ICRS changing at this latitude and of the poorer astrometry of PPMXL south of -30$^{\circ}$.
\item After subtracting the surfaces, bulk motions are smaller and less coherent.
\end{enumerate}

\subsection{Consistency with the ICRS}\label{sec:consistency_with_the_icrs}

We further test the correction volume by considering the data sets which are fundamental to the ICRS. These data sets built the ICRF2 (Fey et al. 2009), the representation of the ICRS. The first data set are 1217 so-called non-VCS (Very Long Baseline Array [VLBA] Calibrator Survey) compact radio sources, among which are the 295 primary ``defining" sources covering the entire celestial sphere. Additionally, 2197 compact radio sources are observed only in VCS sessions north of -45$^{\circ}$ declination.

We collect these sources and their PPMXL proper motions and perform the same correction as above. The average proper motions before and after the correction for our WISE, VCS and non-VCS corrections are shown in Table 1. For reference we also show the same values for the SDSS pipeline proper motions for the set of extragalactic objects described in Section \ref{sec:data} and shown in Figure \ref{fig:color_cut}.

Our correction shows little to no improvement for the proper motion in $\alpha$ for the WISE, VCS and non-VCS data. The most noticeable improvement is in the proper motion in $\delta$ -- a definite trend toward negative values is recentered to $\sim$0 mas yr$^{-1}$. In Table 1 we also compare the PPMXL proper motions of the quasar sources of the \citet{ver2010} catalog before and after the correction. The results are similar to those of the VCS and non-VCS sources.

After this correction, the proper motions of extragalactic sources in PPMXL are similar in magnitude to those of the SDSS proper motion data products.

\begin{figure*}[ht!]
\centering
\includegraphics[height=0.455\textheight]{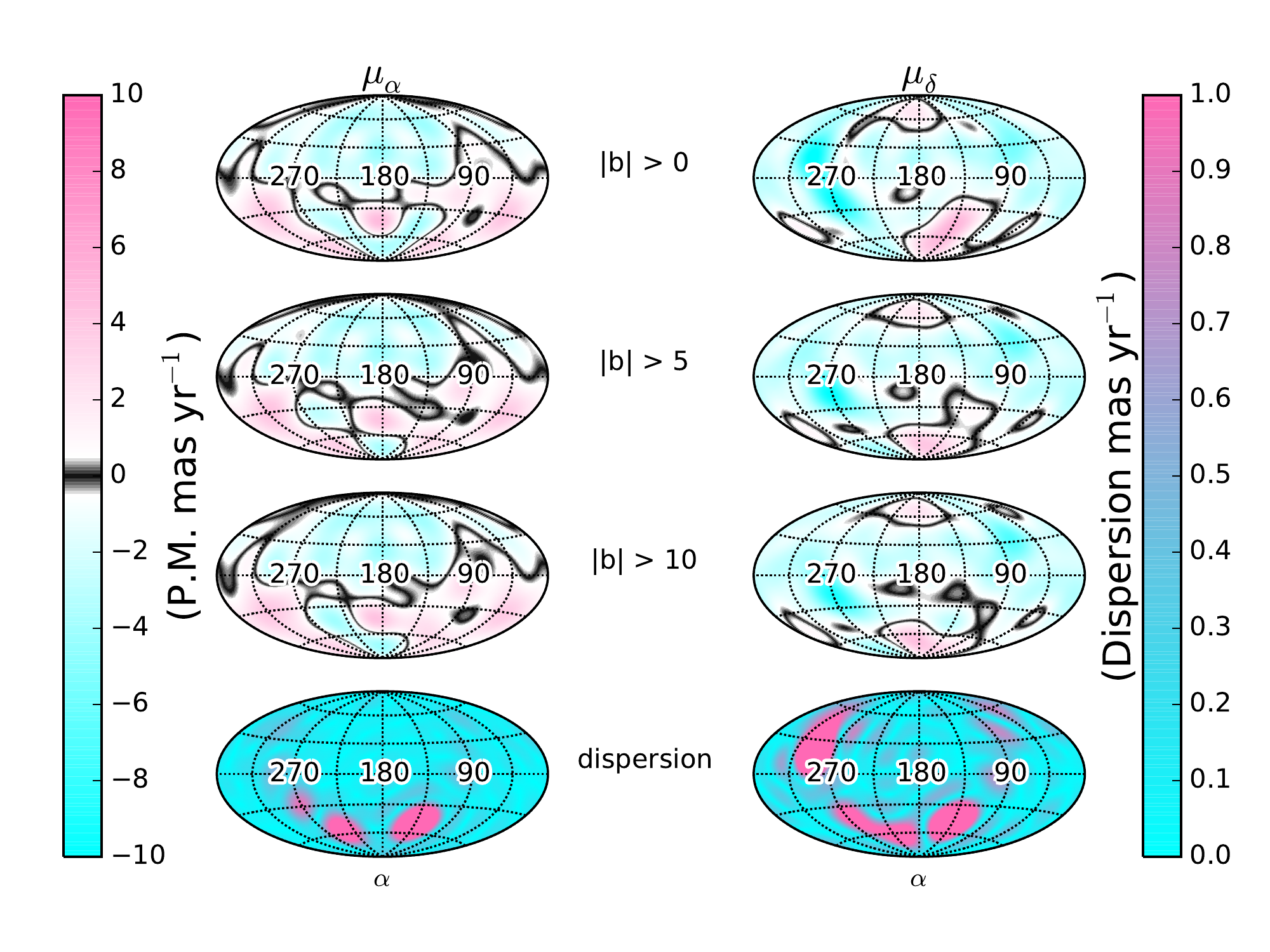}
\caption{The top three panels are three different eighth degree harmonic fits to the proper motions of extragalactic objects at magnitude 15. $<$ J $<$ 15.1. The three fits are performed on three different data sets from which various amounts of the plane are ignored (see center notations). The bottom panel is the dispersion of the three fits and indicates how well the fits agree. The dispersion spikes in the plane of the galaxy, but is relatively low in the halo. This indicates that the fits only differ in the areas where the data is removed, and that the rest of the fits are consistent with each other. In light of this, we fit all of the available data in our work -- instead of ignoring the plane -- and instead warn that the results may be inaccurate in the plane.}
\label{fig:fit_comp}
\end{figure*}

\subsection{Consistency with SDSS Corrective Studies}\label{sec:consistency_with_sdss_corrective_studies}

A similar corrective procedure for the proper motions of quasars in the PPMXL catalog has been carried out by \citet{gra2014}. They produce a corrective 3$^{\circ}$x3$^{\circ}$ grid for the proper motions of quasars identified by the Large sky Area Multi-Object fiber Spectroscopic Telescope (LAMOST; \citealt{cui2012}) and the active galactic nuclei catalog of \citet{ver2010} which have PPMXL proper motions.

Their correction is possibly more accurate on scales consistent with their bin width, however it is incomplete over the full sky and inherently discontinuous. The study is largely constrained to the SDSS footprint, where the majority of known quasars are found, and so only covers about two thirds of the sky. Our work, in contrast, corrects over the full sky, thanks to the 2MASS and WISE all-sky footprints (excepting perhaps low galactic latitudes, see Sections \ref{sec:data} and \ref{sec:eff_plane}). Another advantage to our fits is the magnitude dependence we implement which is only made possible by the statistically large numbers of objects we use.

Comparing our spherical harmonic fits to their binned corrections within the footprint of their study helps to verify our results and lend credence to our correction outside of their footprint. In Figure \ref{fig:grabowski} we plot their corrective surface in $\alpha$ and $\delta$. As noted before in Section \ref{sec:fitting}, we see hot and cold spots on the footprint surface. We then subtract our own (average) spherical harmonic correction at the center of each bin from the corrective surface and plot the residuals below. The most important thing to note is that the hot and cold spots give way to essentially random noise. This implies that the global corrections are similar.

\subsection{The Effect of the Plane}\label{sec:eff_plane}

As shown in Figure \ref{fig:star_vs_gal_sph_hammer}, stellar contamination may affect our sample at low galactic latitudes; there is some concern that the spherical harmonic fits may be corrupted by this contamination. To examine the effect of this possible disk contamination, we perform the spherical harmonic fits described in Section \ref{sec:fitting} on three different data sets: the entire data set, the data set where $|b| > 5^{\circ}$, and the data set where $|b| > 10^{\circ}$.

In Figure \ref{fig:fit_comp} we show the spherical harmonic fits to these three data sets and the dispersion between these three fits. The fits appear quite similar, and the dispersion plot confirms that they are well matched in the off-plane sections of the sky. The dispersion, measuring the mismatch of the three fits, is highest in the plane and relatively small in the halo.

The three fits agree with each other to within 1.5 mas yr$^{-1}$ for any given point more than $\sim$20$^{\circ}$ from the plane. Figure \ref{fig:quiver} is remade for the fits excluding the data $|b| < 5^{\circ}$, the data $|b| < 10^{\circ}$ and compared to the original figure (not shown here). The only noticeable difference is that the plane is more chaotic in the fits eliding planar data.

Since the fits do not appear to differ much outside the plane, we opt to fit all of the data without excising the planar regions from the fit. Instead we warn that the fits may be inaccurate in the plane of the Galaxy and up to 20$^{\circ}$ from the plane.

\section{Conclusion}\label{sec:discussion}

We have collected a sample of extragalactic objects using WISE and 2MASS colors as a discriminator. The PPMXL proper motions for these objects are then fitted to combinations of spherical harmonics in seven different magnitude slices. Considering that the intrinsic proper motions of these objects are negligible, we assume that the proper motions of objects in these slices indicate an offset from zero in the PPMXL proper motion catalog.

Interpolating between these slices generates a magnitude dependent correction which may be subtracted from any given point in the PPMXL database with a J magnitude measurement to produce a new, corrected proper motion estimate. Note that this correction may be erroneous when considering extreme reddening regions such as the Galactic plane. We would suggest using this correction only at absolute latitudes higher than 20$^{\circ}$.

The PPMXL catalog, along with this correction, will be superseded by the astrometric data from the Gaia satellite \citep{per2001} which will become the optical representation of the ICRS. However, for objects fainter than the Gaia limit, a technique similar to this may be used to refer their relative proper motions to the ICRS.

We make available code\footnote{available at time of publication of this preprint at: \url{https://github.com/johnjvickers/ppmxl_correction}} for performing such a correction and append new columns to the PPMXL database at the German Astrophysical Virtual Observatory\footnote{http://www.g-vo.org} which have been corrected in this fashion.

\acknowledgments

We thank Dr. Markus Demleitner for his invaluable assistance with some of the technical aspects of this work and his aid in putting the proper motion corrections into the German Astrophysical Virtual Observatory PPMXL database.

We thank Dr. Annie Robin for her comments on an earlier draft which helped improve the clarity of the figures and the robustness of the result.

We also thank Drs Bertrand Goldman, Ciprian Berghea, and Jeff Carlin for helpful discussion and direction.

We thank the anonymous referee for their numerous comments and notable precision.

We thank the developers and maintainers of the following software libraries which were used in this work: Topcat \citep{tay2005}, Aladin \citep{bon2000}, NumPy, SciPy, AstroPy \citep{ast2013}, astroML \citep{van2014}, matplotlib and Python.

This publication makes use of data products from the Wide-field Infrared Survey Explorer, which is a joint project of the University of California, Los Angeles, and the Jet Propulsion Laboratory/California Institute of Technology, funded by the National Aeronautics and Space Administration.

This publication makes use of data products from the Two Micron All Sky Survey, which is a joint project of the University of Massachusetts and the Infrared Processing and Analysis Center/California Institute of Technology, funded by the National Aeronautics and Space Administration and the National Science Foundation.

This work was supported by the Marie Curie Initial Training Networks grant number PITN-GA-2010-264895 ITN ``Gaia Research for European Astronomy Training" and by Sonderforschungsbereich SFB 881 ``The Milky Way System" (subproject A2 and A3)
of the German Research Foundation (DFG). J.J.V. is a fellow at the International Max Planck Research School for Astronomy and Cosmic Physics at the University of Heidelberg; JJV also acknowledges the support of a LAMOST fellowship and the National Key Basic Research Program of China grant number 2014CB845700. This work is also supported by the CAS One Hundred Talent Fund and from NSFC grants 11173002 and 11333003, and the Strategic Priority Research Program The Emergence of Cosmological Structures of the Chinese Academy of Sciences, Grant No. XDB09000000.

\end{document}